\documentclass[aps,prl,
twocolumn,
superscriptaddress]{revtex4}  
\usepackage{graphicx}         
\usepackage{ragged2e}       
\usepackage{amssymb,amsmath,xcolor}
\usepackage[caption=false]{subfig}
\usepackage{xcolor}
\usepackage[export]{adjustbox}
\usepackage{rotating}

\begin{document}

\title{Absorption and opacity threshold for a thin foil \\in a strong circularly polarized laser field}
\author{E.G.~Gelfer}\email{egelfer@gmail.com}\affiliation{ELI Beamlines, Institute of Physics of the ASCR, v.v.i., Dolni Brezany, Czech Republic}\affiliation{National Research Nuclear University MEPhI, Moscow, Russia}
\author{A.M.~Fedotov}\affiliation{National Research Nuclear University MEPhI, Moscow, Russia}
\author{O.~Klimo}\affiliation{ELI Beamlines, Institute of Physics of the ASCR, v.v.i., Dolni Brezany, Czech Republic}\affiliation{FNSPE, Czech Technical University in Prague, Czech Republic}
\author{S.~Weber}\affiliation{ELI Beamlines, Institute of Physics of the ASCR, v.v.i., Dolni Brezany, Czech Republic}\affiliation{School of Science, Xi'an Jiaotong University, Xi'an 710049, China}


\begin{abstract}
We show that a commonly accepted transparency threshold for a thin foil in a strong circularly polarized normally incident laser pulse needs a refinement. We present a new analytical model, which correctly accounts for laser absorption. The refined threshold is determined not solely by the laser amplitude, but other parameters are equally or even more important. Our predictions are in a perfect agreement with PIC simulations. The refined criterion is crucial for configuring laser plasma experiments in the high field domain. Besides, an opaque foil steepens the pulse front, this can be important for numerous applications.
\end{abstract}
\maketitle


A key attribute specifying the laser-matter interaction regime is transparency or opaqueness of a target to an incident laser pulse \cite{mourou2006,macchibook}. Precise formulation of a relevant discriminating criterion is  crucially important for many modern applications, including laser contrast enhancement by plasma shutters \cite{reed2009,palaniyappan2012}, laser frequency upshift by relativistic electron mirrors \cite{kiefer2013}, plasma based polarizers \cite{stark2015}, generation of gamma-rays \cite{brady2014,chang2017}, attosecond pulses \cite{ji2011} and bright neutron bunches \cite{roth2013} in laser plasmas, as well as for laser plasma ion acceleration \cite{esirkepov2004,macchi2009,gonoskov2009,yin2011,mackenroth2016,matys2019}. The latter can be applied for a wide range of purposes (see the review \cite{macchi2013}) from fast ignition in fusion targets \cite{naumova2009} to cancer treatment \cite{bulanov2004}. As radiation pressure acceleration of ions is most efficient when the target density is kept slightly above the threshold \cite{esirkepov2004,macchi2013}, it is reasonable to optimize the laser pulse temporal profile to maintain such a condition as long as possible \cite{bulanov2012} (though other mechanisms \cite{gonoskov2009,mackenroth2016} may favor the foil density slightly below the threshold). In contrast, transparent targets are preferable for observation and detailed studies of an impact of radiation friction on plasma dynamics \cite{tamburini2010,trapping,longfield} or strong field QED effects \cite{dipiazza2012}, as otherwise fewer electrons can ever reach the focal region to probe the strongest field. Such studies are planned with the next generation laser facilities under construction \cite{ELI, ELINP, Apollon, guo2018}. 

As is well known \cite{chen}, a moderately intense laser pulse (of dimensionless field strength $a_0\equiv eE_0/m\omega c \lesssim 1$)
can penetrate through a thick plasma target (of thickness $d\gg\lambda$, where $\lambda$ is the laser carrier wavelength) only if the unperturbed target density $n_0$ is less than the critical value $n_{cr}=m\omega^2/4\pi e^2$. Here $E_0$ is the electric field strength amplitude, $\omega$ is the laser carrier frequency, $m$ and $-e$ are electron mass and charge. For a strong pulse, $a_0\gg1$, this condition is modified to $n_0\lesssim a_0n_c$ \cite{ap1956,kaw1970} due to the effect of relativistic self-induced transparency (RSIT), see further discussion on the elaboration of this criterion in   \cite{cattani2000,goloviznin2000,weng2012,siminos2012,siminos2017,ji2018}.

Here we study an opposite case of a thin foil target ($d\lesssim\lambda$), assuming that the laser pulse has ultrarelativistic intensity ($a_0\gg 1$), is circularly polarized and is incident normally (the latter simplifications are imposed to minimize electrons heating). For such a case the transparency threshold condition obtained in \cite{vshivkov1998} (see also \cite{macchi2009}) involves the dimensionless electron areal density $\sigma_0=(n_0/n_{cr})\cdot(\omega d/c)$ of the foil and can be cast to the form
\begin{equation}\label{tc}
\sigma_0<\sigma_0^{th}=2a_0.
\end{equation}

\begin{figure*}[t!]
\subfloat{\includegraphics[width=0.45\linewidth]{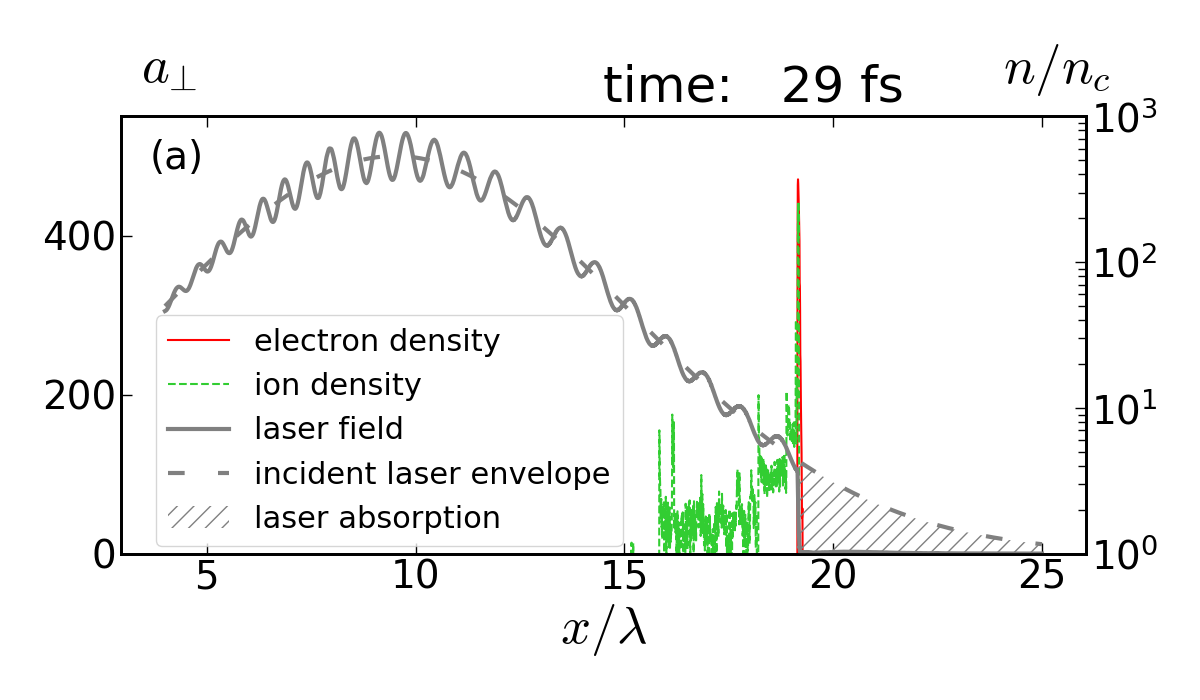}}
\subfloat{\includegraphics[width=0.45\linewidth]{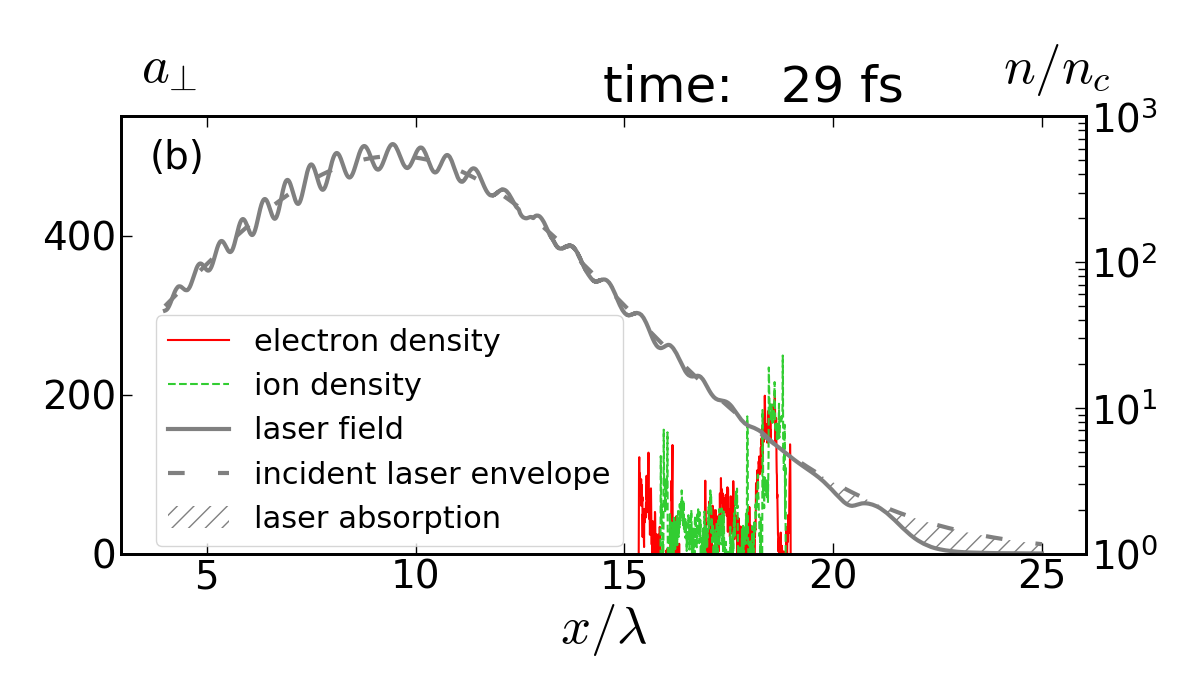}}
\caption{\label{fig1} 1D PIC simulation results for the interaction of a strong circularly polarized Gaussian laser pulse ($\lambda=1\mu$m, $a_0=500$, FWHM duration $30$fs) with a thin hydrogen foil (thickness $d=0.1\lambda$). (a): foil electron density $n_0=140 n_c$ (areal density $\sigma_0\approx 88$) -- the foil is opaque; (b): foil electron density $n_0=80 n_c$ (areal density $\sigma_0\approx 50$) -- the foil is transparent.}
\end{figure*}

However, according to the results of PIC simulations \footnote{All the simulations in the paper are performed with the code SMILEI \cite{SMILEI}. The laser pulse propagates from left to right, the target is initially located at $x_0=15\lambda$ having a rectangular density profile $n(x)=n_0\theta(x-x_0)\theta(x_0+d-x)$, where $\theta(x)$ is the Heaviside step function. The cell size is $\lambda/200$, the time step satisfies the the Courant-Friedrichs-Lewy condition. The boundaries are absorptive. Each cell contains $800$ particles of each type and $n_0/n_c$ particles of each type in 1D and 2D case, respectively.}, the threshold value in Eq.~(\ref{tc}) needs a refinement if $a_0$ is a few hundred or higher (see the precise condition (\ref{vcond}) below). For example, a foil with $\sigma_0\approx 88$ can still remain opaque to a laser pulse even for $a_0=500$ (see Fig.~\ref{fig1}a), i.e. for more than an order of magnitude higher value than prescribed by the threshold (\ref{tc}). 
The reason is that in a derivation of the condition (\ref{tc}) it is assumed that laser absorption is negligible, so that an incident laser pulse is totally reflected by an opaque target. But it turns out that for high enough values of $a_0$ the absorption can become crucially important, thus setting up a lower areal density limit for RSIT, which then depends more on a laser pulse duration and envelop shape than on its amplitude. 

\begin{figure}[h!]
\subfloat{\includegraphics[width=0.9\linewidth]{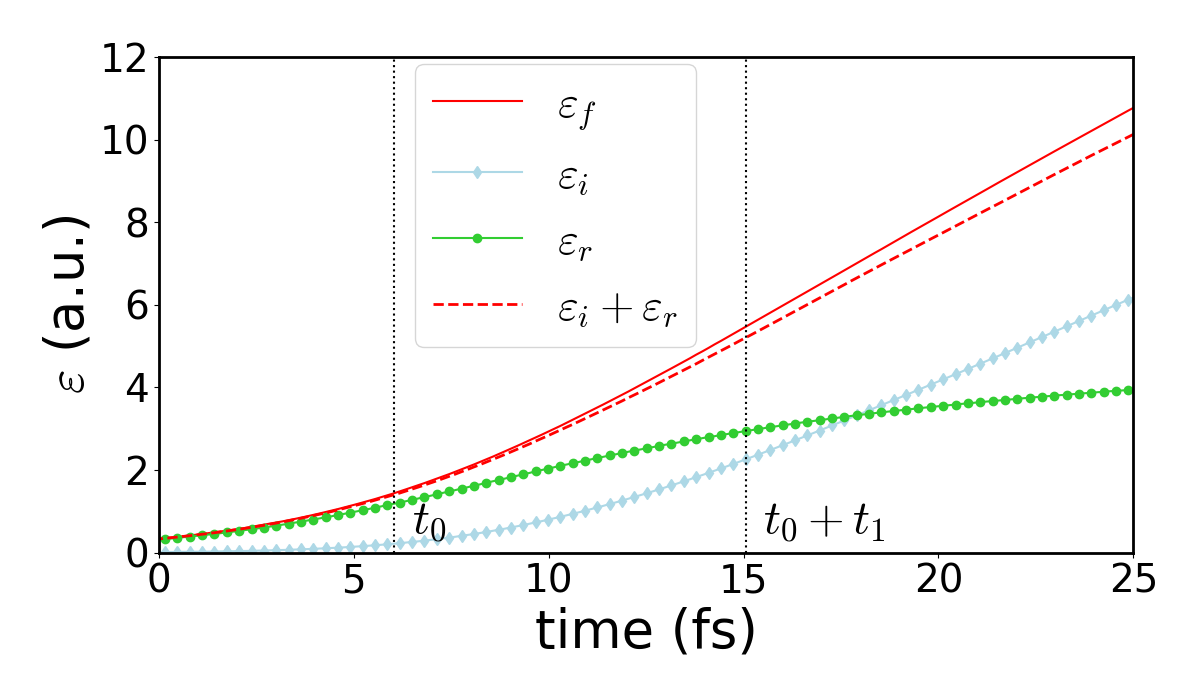}}
\caption{\label{fig1c} Time dependencies of the laser front (hatched in Fig.1a) energy ($\varepsilon_f$), the ions energy ($\varepsilon_i$) and the reflected energy ($\varepsilon_r$) for the simulation presented in Fig. 1a.}
\end{figure}

Let us illuminate the role of laser absorption using a 1D laser--foil interaction model with a due consideration of the laser pulse temporal profile. Denote by $a(t)$ the value of the dimensionless laser field strength at time $t$ at the foil position $x_t$. Upon the incidence of a laser pulse on a foil $a(t)$ is gradually growing from zero onward. Let $t_0$ be the least moment when $a(t_0)=\sigma_0/2$. Then, according to Eq. (\ref{tc}), while $t<t_0$ the laser pulse front is totally reflected by the foil, as shown in Fig.~\ref{fig1c}. Hence the field behind the foil vanishes, $\mathbf{E}(t,x>x_t)=0$, and the laser profile eventually takes a step-like form similar to the one shown in  Fig.~\ref{fig1} (a). Such a profile creates a strong longitudinal ponderomotive force, which efficiently accelerates the electrons forward. Electrons in turn pull the ions behind themselves, therefore the whole plasma gets accelerated in front of the laser pulse \cite{SM}. For $t>t_0$ the reflection alone cannot provide the opacity of the foil and, assuming the transition to transparency is procrastinated, upon the further growth of $a(t)$ the rest of the pulse bulk is being partially reflected and partially absorbed. The absorbed energy is eventually transferred to the ions, enforcing their acceleration \footnote{Note that since the energy is eventually almost totally transferred form electrons to ions, the specific mechanism of energy absorption by electrons is not important for determining the opacity threshold.}. Energy and momentum transfer to the ions reduces the momentum flux that electrons should reflect backwards to maintain the foil opaque. We are especially interested in an initial stage of the process when the ions remain mildly relativistic. To refine the threshold, it is enough to thoroughly balance the energy of the pulse bulk with the sum of the energies of the reflected wave and those transferred to the ions (the energy gained by electrons is much smaller and can be neglected).

Initially, as long as the ions are slow ($v_i\ll c$), they lag behind the electrons, hence do not overlap with them and the dimensionless charge separation field can be estimated as $\sigma_0$. At this stage the longitudinal component of the ions 4-velocity is $u_i\sim \mu Z \sigma_0\omega(t-t_0)/A$, where $A$ and $Z$ are the ion mass and charge numbers, $\mu$ is the electron-to-proton mass ratio. Importantly, it is taken into account that substantial ion acceleration starts only together with laser absorption. The ions become mildly relativistic, $u_i\sim 1$, at $t=t_0+t_1$, where
\begin{equation}\label{t1}
t_1=\frac{1}{\omega}\frac{A}{Z\mu\sigma_0}
\end{equation}
is the required duration of their acceleration. The total energy of the ions at this moment can be estimated as 
\begin{equation}\label{ei}
\varepsilon_i(t_1)\sim \frac{A\sigma_0}{Z\mu}\frac{n_{cr} S c }{\omega} m_e c^2,
\end{equation}
where $S$ is the laser pulse cross section. According to \cite{vshivkov1998}, the amplitude of the reflected wave for $t_0<t<t_0+t_1$ is $a_r=\sigma_0/2$ \footnote{See the discussion after Eqs.~(48) and (49) in \cite{vshivkov1998}. 
}. Therefore the reflected energy during this time interval  can be estimated as 
\begin{equation}\label{er}
\varepsilon_r(t_1)\sim \frac{\sigma_0^2t_1 }{4} n_{cr}S m_ec^3\sim\frac{\varepsilon_i(t_1)}{4}.
\end{equation}
Note that $\varepsilon_r$ approaches saturation for $t>t_0+t_1$, see Fig.~\ref{fig1c}. This is natural as the foil becomes relativistic, so that the intensity of the wave reflected from the relativistic mirror is suppressed due to the Doppler effect. 

Next consider the energy of a part of the laser pulse front having touched the foil by the moment $t_1$ (for the sake of clarity in Figs.~\ref{fig1} (a) and (b) an imaginary envelope for free propagation of the pulse in the absence of the target is plotted dashed and its corresponding part is hatched). For brevity, in what follows let us call it just the laser front energy. Let us write the initial envelope as $a(\varphi)=a_0 g(\varphi/\omega T)$, where $\varphi$ is the phase, $T$ is the characteristic duration and $g$ is the dimensionless profile function of the order of unity decaying at $|\varphi|\gg \omega T$. Then the energy stored in the front of the laser pulse during $t_0<t<t_0+t_1$ can be estimated by 
\begin{equation}\label{ef0}
\varepsilon_f(t_1)\sim a_0^2n_{cr} S m_e c^3\int\limits_{t_0}^{t_0+t_1} g^2(t/T)dt.
\end{equation}
The integral encountering in Eq.~(\ref{ef0}) can be approximated as 
\begin{equation}\label{appr}
\int_{X_0}^{X_0+X_1} f(\xi)d\xi\approx\frac{ f^2(X_0)}{ f'(X_0)}\left[e^{X_1 f'(X_0)/f(X_0)}-1\right],
\end{equation}
if $|f''(X_0)f(X_0)/(f'(X_0))^2-1|\ll 1$. 

If the foil remains opaque for $t\sim t_0+t_1$ then, due to absence of transmission, this energy should be eventually partially reflected and in the rest part entirely absorbed by the ions and hence should be equal to the sum of (\ref{ei}) and (\ref{er}), see Fig.~\ref{fig1c}. 
Hence, by equating the sum of (\ref{ei}) and (\ref{er}) to (\ref{ef0}) and taking into account (\ref{appr}) and that
\begin{equation}\label{asigma}
a_0g(t_0/T)=\sigma_0/2,
\end{equation}
we arrive at
\begin{equation}\label{zeta}
e^\zeta-1\approx 5\zeta,
\end{equation}
where $\zeta=2A|g'(t_0/T)|/(Z\mu\sigma_0\omega Tg(t_0/T))$. Equation (\ref{zeta}) has an approximate solution $\zeta\approx 2.66$, and finally the refined opacity threshold can be formulated in the form
\begin{equation}\label{opcond}
\mu\omega T\sigma_0\frac{Z}{A}\frac{ g(t_0/T)}{| g'(t_0/T)|}\approx 0.75,
\end{equation}
where $t_0$ is determined by (\ref{asigma}). Note that our threshold value is singled out by the  (violated in a deeply overcritical case) assumption of complete charge separation in estimating $t_1$.

\begin{figure*}[ht!]
\subfloat{\includegraphics[width=0.4\linewidth]{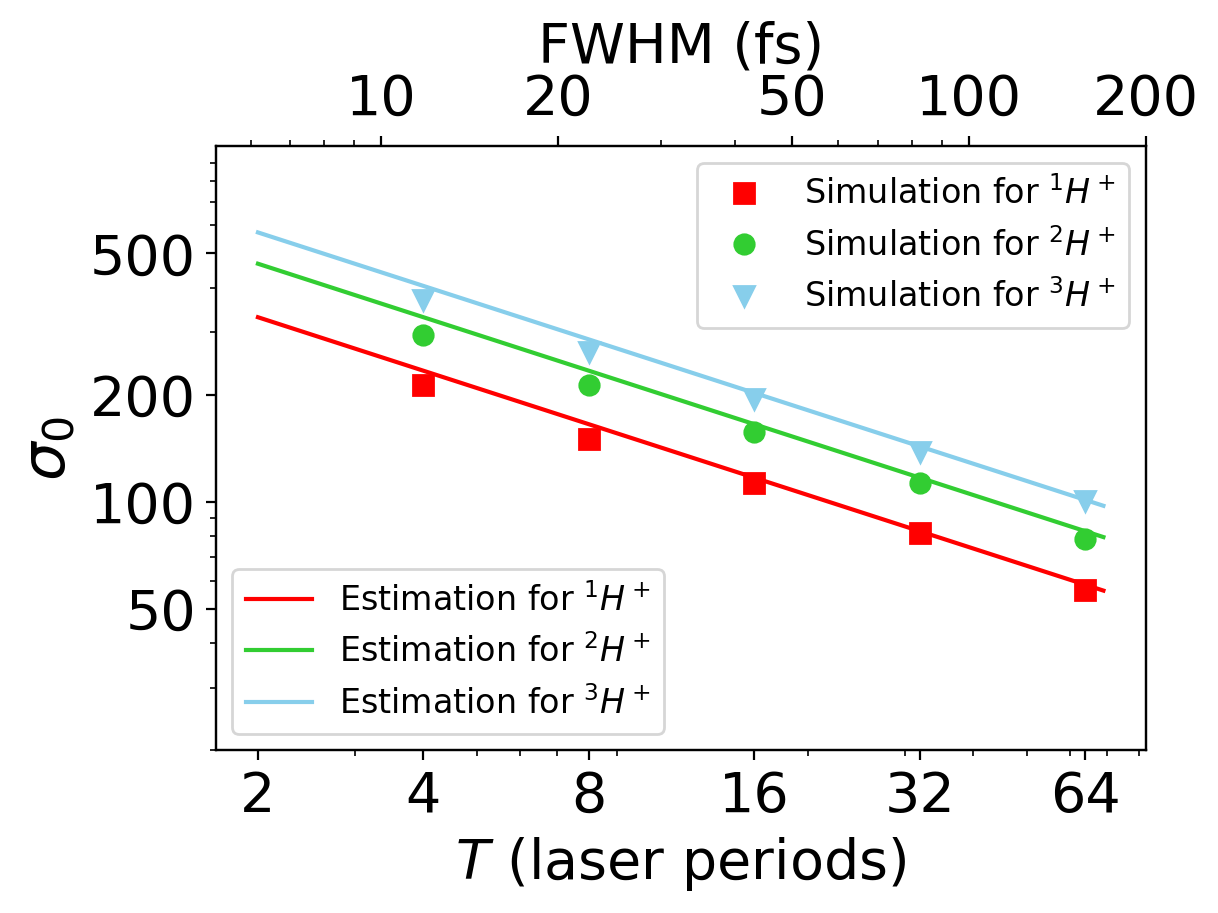}}\qquad
\subfloat{\includegraphics[width=0.4\linewidth]{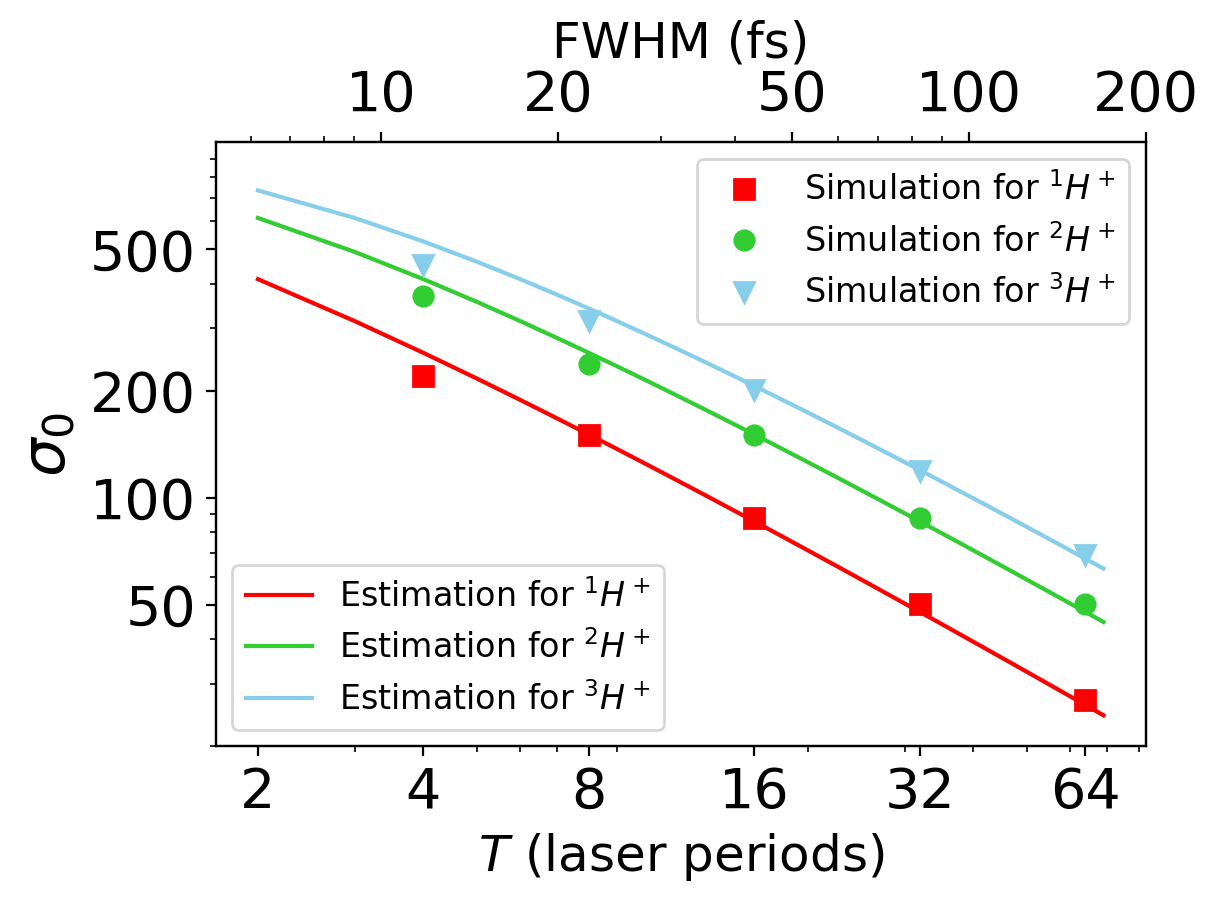}}
\caption{\label{fig3} Estimations vs 1D PIC simulation results for transparency threshold areal density for different pulse durations and ion charge-to-mass ratios. Dimensionless field amplitude $a_0=500$, laser carrier wavelength $\lambda=1\mu$m, foil thickness $d=0.1\lambda$. Left panel: linear pulse profile, right panel: Gaussian pulse profile.}
\end{figure*}

Let us briefly discuss the range of applicability of the condition (\ref{opcond}). The approximation (\ref{appr}) can be used only for $|t_0|\gg T$, as otherwise reflection dominates over the absorption and the usual transparency condition Eq. (\ref{tc}) remains valid. It means that $\sigma_0\ll a_0$ and hence
\begin{equation}\label{vcond}
\quad a_0\gg \frac{A}{\omega TZ\mu}.
\end{equation}
It is also worth to emphasize that our approach is reasonable only for a normally incident circularly polarized pulse as otherwise a double layer consisting of an electron spike followed by the ions [see Fig.~\ref{fig1} (a)] is rapidly messed up due to electron heating \cite{macchi2005,klimo2008}.


In order to compare Eq.~(\ref{opcond}) with numerical simulations, let us illustrate it with the two particular examples of Gaussian and linear profiles. Namely, if the function $g(\xi)$ is Gaussian, $g(\xi)=e^{-\xi^2}$, then the threshold areal density $\sigma^G_{th}$ is determined by 
\begin{equation}\label{sigmag}
\frac{\sigma^G_{th}}{\sqrt{\ln(2a_0/\sigma^G_{th})}}\approx 1.5\frac{A}{\omega TZ\mu }.
\end{equation}
Note that the refined threshold depends on the laser field amplitude $a_0$ much weaker than on the remaining parameters (pulse duration, ion mass and charge number). For our second example of a linear profile [$g(\xi)=0$ for $\xi<0$, $g(\xi)=\xi$ for $0<\xi<1$, and $g(\xi)=1$ for $\xi>1$] the approximation (\ref{appr}) is invalid, but the integral in (\ref{ef0}) is easily evaluated directly, so that instead of (\ref{opcond}) we arrive at
\begin{equation}\label{sigmal}
\sigma_{th}^L\approx\sqrt{\frac{1.5a_0}{\omega T}\frac{A}{Z\mu}}.
\end{equation}
One can observe that here the refined threshold is fully determined by the envelope slope rather than by laser pulse amplitude or duration separately.

The obtained estimations Eqs. (\ref{sigmag}) -- (\ref{sigmal}) for the areal threshold densities are compared to the values obtained by 1D PIC simulations in Fig.~\ref{fig3}. Namely, we performed a set of simulations for several values of pulse duration and for three types of ions: hydrogen ($Z/A=1$), deuterium ($Z/A=0.5$), and tritium ($Z/A=0.33$). The field strength amplitude was taken $a_0=500$, but for a Gaussian pulse we have also checked that the results remained almost unchanged as $a_0$ was increased up to $1000$ and $1500$. We used targets of fixed thickness $d=0.1\lambda$ but varied the density, and considered the foil as opaque, if the intensity of the laser field to the right from the foil is negligible compared with the intensity to the left (compare, e.g. figures \ref{fig1} (a) and (b)). One can observe a perfect agreement between the theory and simulations, though, as expected, the simulation results start to deviate from the model when $\sigma_{th}$ is so high that approaches $a_0$.


\begin{figure*}[ht!]
\subfloat{\includegraphics[width=0.4\linewidth,valign=t]{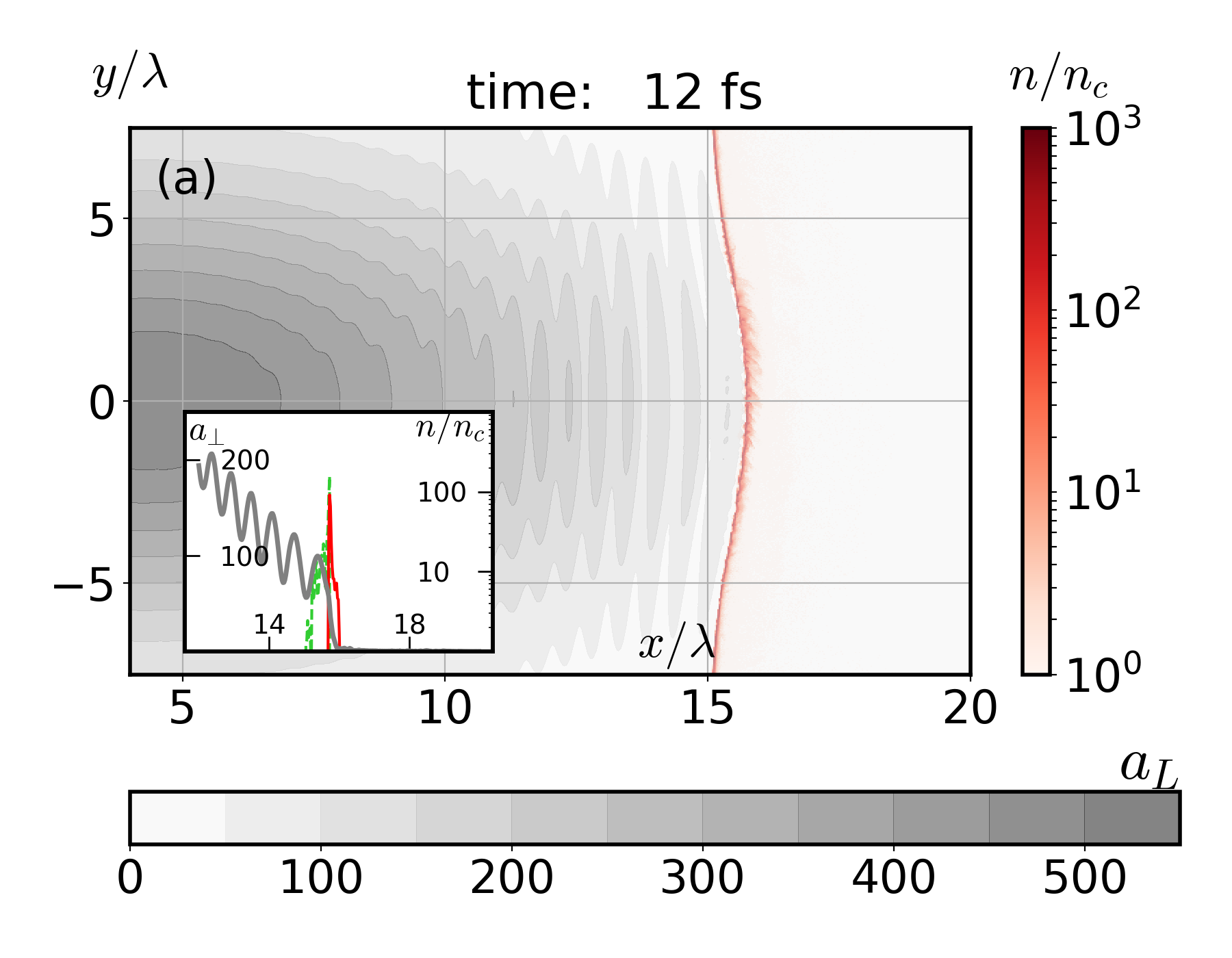}}
\subfloat{\includegraphics[width=0.4\linewidth,valign=t]{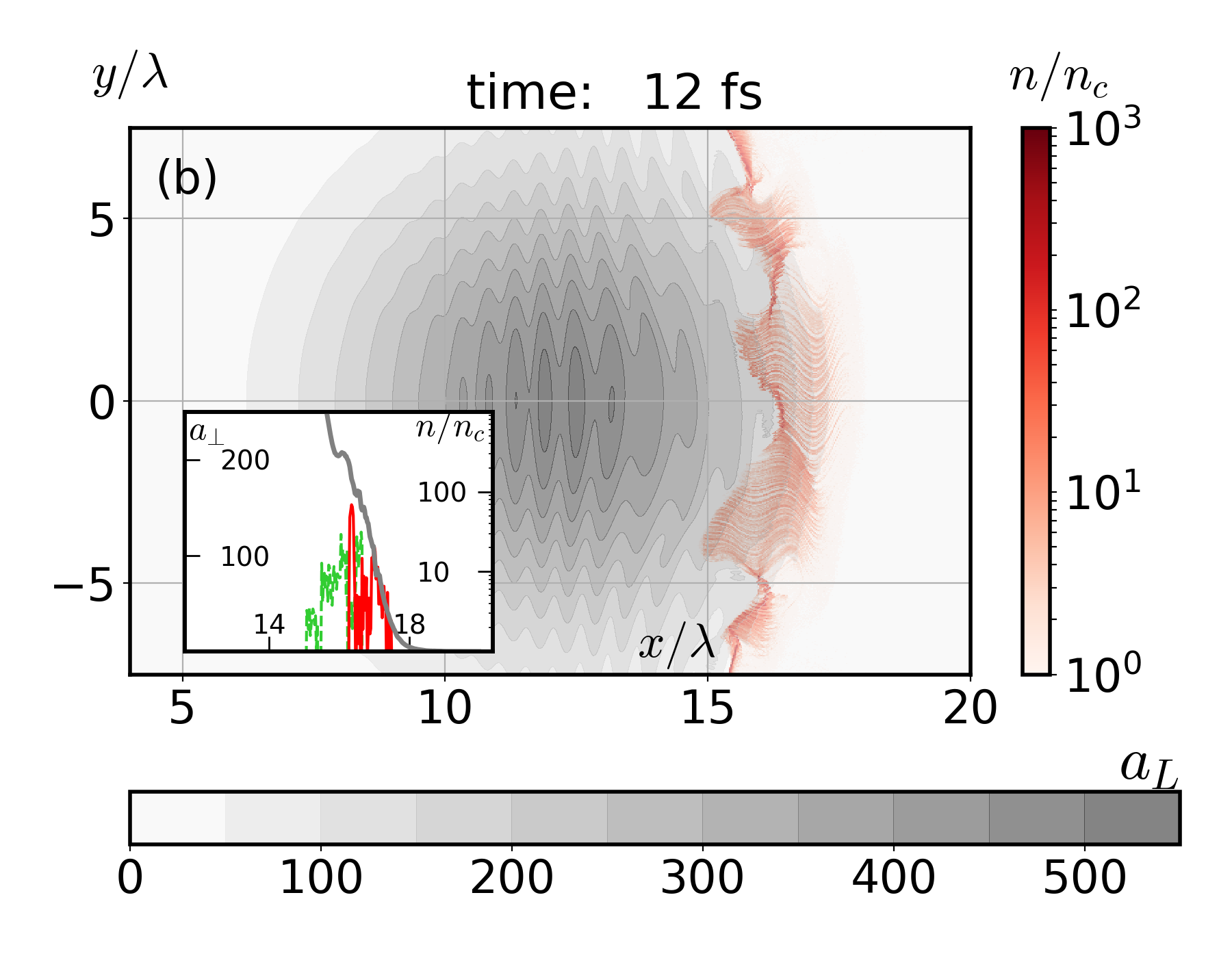}}\\
\subfloat{\includegraphics[width=0.4\linewidth,valign=t]{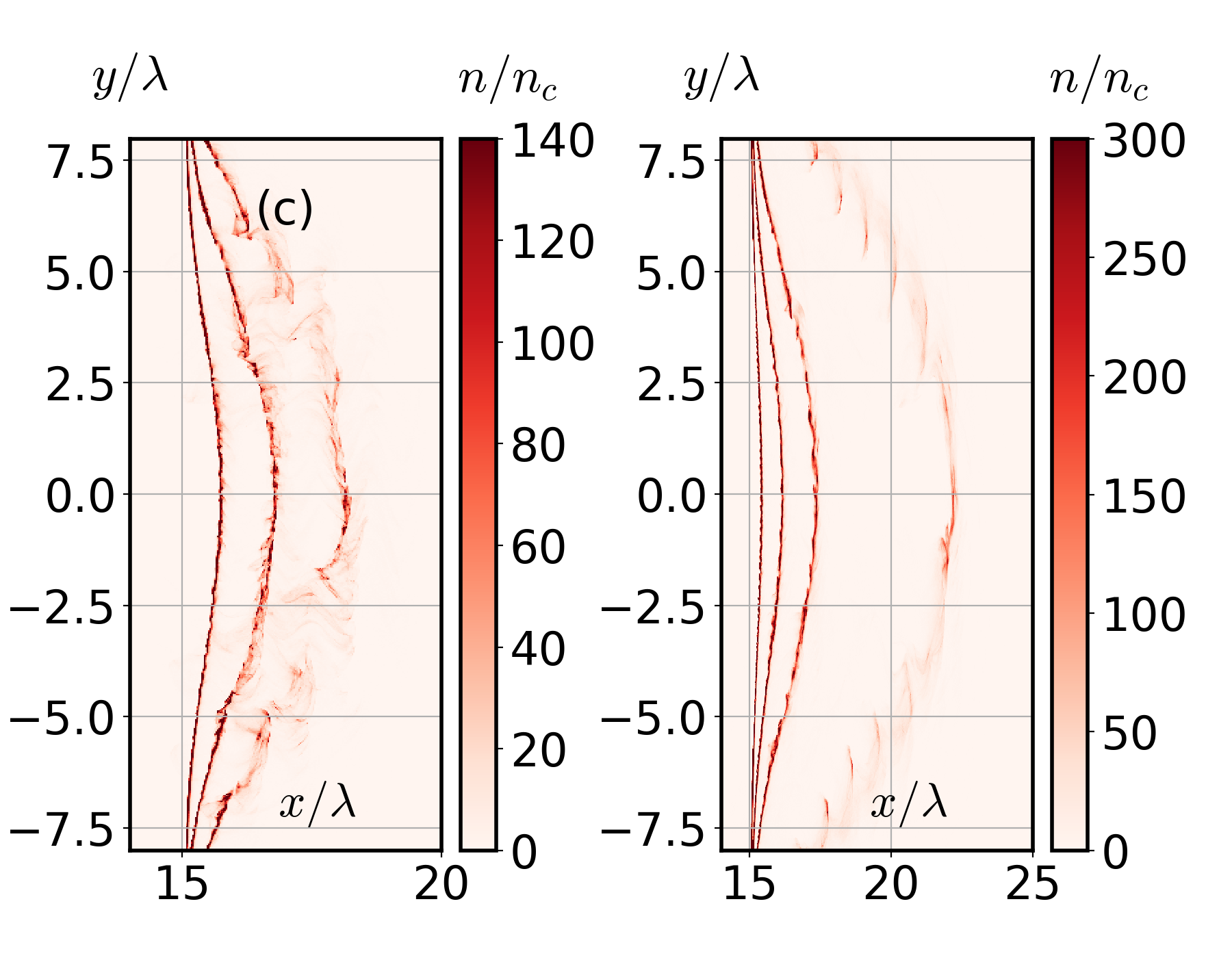}}
\subfloat{\includegraphics[width=0.4\linewidth,valign=t]{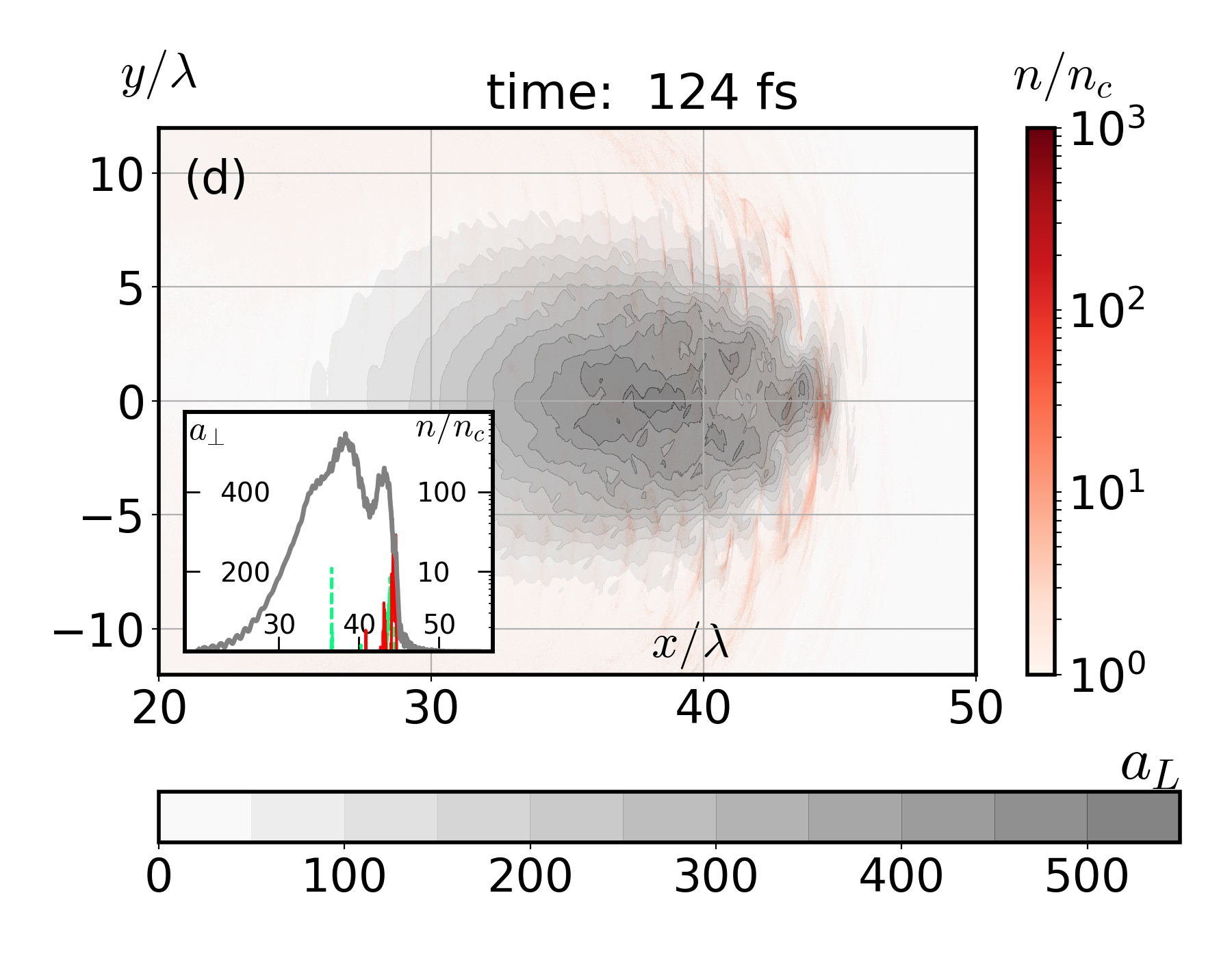}}
\caption{\label{fig4} 2D PIC simulation results. $a_0=500$, transverse waist radius $w=6\lambda$, $\lambda =1\mu m$, $d=0.1\lambda$, $H^+$ ions.  (a), (b), (d): electron density (red, continuous color bar), and laser field (grey, discrete color bar) distributions for laser pulse FWHM 30 fs [(a) and (d)] and 15 fs [(b)], target density $n_0=140 n_c$ [(a) and (b)] and $n_0=300 n_c$ [(d)]; insets (the same legends as in Fig.~\ref{fig1}): a sectional view at $y=0$ . (c): electron density distributions at subsequent moments of time for different target densities. Left: $n_0=140 n_c$ and $t=12$ fs, 18 fs, 26 fs; right: $n_0=300 n_c$ and $t=12$ fs, 18 fs, 26 fs, 46 fs.}
\end{figure*}

The main additional effect arising in a 2D case is the Rayleigh-Taylor instability (RTI) of a thin foil irradiated by a strong laser pulse. In the case of relativistic motion of the foil its growth time is  \cite{pegoraro_prl2007} 
\begin{equation}\label{taurt}
t_{RT}\sim\frac{1}{\omega}\sqrt{\frac{\mu\sigma_0}{A}}\frac{1}{k^{3/2}},
\end{equation}
where $k$ is the dimensionless wave vector of a perturbation and Eq. (\ref{asigma}) is taken into account. 

To check reliability of our estimations in 2D we performed 2D PIC simulations for a pulse with Gaussian profiles in both longitudinal and transverse directions (see Fig.~\ref{fig4}). It turns out that our estimate (\ref{sigmag}) for the transparency threshold areal density still remains valid. This can be confirmed by comparing Figs~\ref{fig4} (a) and (b), where the interaction of laser pulses of different durations but of the same amplitude with identical targets is displayed. The duration of the laser pulse in Fig.~\ref{fig4} (a) according to the condition (\ref{sigmag}) corresponds to opacity, while the pulse in Fig.~\ref{fig4} (b) is taken twice shorter, resulting in target transparency. 

At later times RTI distorts the target and eventually destroys it, see Fig.~\ref{fig4} (c), therefore it is meaningless to discuss its transparency for times $t\gg t_{RT}$. However, since $t_{RT}$ is proportional to $\sqrt{\sigma_0}$, denser targets survive under RTI longer, see Fig.~\ref{fig4} (c) and also \cite{SM}. Besides, similarly to the 1D case, target opaqueness leads to the steepening of a laser front, see Fig.~\ref{fig4} (d), and after the destruction of the target the resulting steepened pulse with an increased femtosecond level contrast can be used for applications \cite{reed2009,wang2011,palaniyappan2012,wei2017}.


To conclude, we have demonstrated that a commonly accepted threshold for opaqueness of a thin foil to a strong circularly polarized laser pulse needs a revision. It is shown that such a refinement is due to laser absorption not properly taken into account in previous studies. As a consequence, with shorter laser pulses RSIT can be achieved for smaller areal density of a target. Interestingly, it turns out that for Gaussian pulses the refined threshold areal density is almost independent of the field amplitude, depending only on pulse duration and the ions charge-to-mass ratio. Our findings are in excellent agreement with 1D PIC simulations. They are in agreement with 2D simulations as well, though in a 2D case the effect is strongly distorted by the Rayleigh-Taylor instability.

Moreover, if a front part of the pulse gets totally absorbed like in Fig.~\ref{fig1} (a), then the target acts as a plasma shutter \cite{reed2009,wang2011,palaniyappan2012,wei2017} steepening the pulse front and increasing laser contrast on a femtosecond level. This can be important for a wide range of applications, including high harmonic generation \cite{thaury2007,behmke2011} and laser ion acceleration. In the latter case high contrast enhances the energy \cite{ceccotti2007} and suppresses the divergence \cite{green2014,fang2016} of the accelerated ion beams, it is also crucial for such highly efficient ion acceleration mechanisms as radiation pressure acceleration \cite{esirkepov2004,robinson2008,matys2019} and breakout afterburner \cite{yin2011}. Though the effect reported here reveals in thick targets as well, see e.g. \cite{ji2018}, its proper description in such a case requires further studies.

\begin{acknowledgments}
We are grateful to M. Grech and Y.J. Gu for valuable discussions. 
The results of the project LQ1606  were obtained with the financial support of the Ministry of Education, Youth and Sports of the Czech Republic as part of targeted support from the National Programme of Sustainability II.
The research was performed using the code SMILEI \cite{SMILEI} and the resources of the ELI Beamlines Eclipse cluster, and was partially supported by the MEPhI Academic Excellence Project (Contract No.~\texttt{\detokenize{02.a03.21.0005}}), Russian Foundation for Basic Research (Grant \texttt{\detokenize{19-02-00643}}), Czech Science Foundation Project No. 18-09560S, projects ELITAS (ELI Tools for Advanced Simulation) \texttt{\detokenize{CZ.02.1.01/0.0/0.0/16_013/0001793}}, ADONIS (Advanced research using high intensity laser produced photons and particles) 
\texttt{\detokenize{CZ.02.1.01/0.0/0.0/16_019/0000789}} and HiFI (High-Field Initiative) \texttt{\detokenize{CZ.02.1.01/0.0/0.0/15003/0000449}} from European Regional Development Fund. 
\end{acknowledgments}



\begin{thebibliography}{99}

\bibitem{mourou2006}G. A. Mourou, T. Tajima, and S. V. Bulanov, Rev. Mod. Phys. {\bf 78}, 309 (2006).
\bibitem{macchibook}A. Macchi, {\it A Superintense Laser--Plasma Interaction Theory Primer} (Springer, 2013).


\bibitem{reed2009} S. A. Reed, T. Matsuoka, S. Bulanov, M. Tampo, V. Chvykov, G. Kalintchenko, P. Rousseau, V. Yanovsky, R. Kodama, D. W. Litzenberg {\it et. al.}, Appl. Phys. Lett. {\bf 94}, 201117 (2009).
\bibitem{palaniyappan2012} S. Palaniyappan, B. M. Hegelich, H.-C. Wu, D. Jung, D. C. Gautier, L. Yin, B. J. Albright, R. P. Johnson, T. Shimada, S. Letzring {\it et. al.}, Nature Physics {\bf 8}, 763-–769 (2012).


\bibitem{kiefer2013}D. Kiefer, M. Yeung, T. Dzelzainis, P. S. Foster, S. G. Rykovanov, C. L. S. Lewis, R. S. Marjoribanks, H. Ruhl, D. Habs, J. Schreiber {\it et. al.},  Nat. Comms. {\bf  4}, 1763 (2013).

\bibitem{stark2015}D. J. Stark, C. Bhattacharjee, A. V. Arefiev, T. Toncian, R. D. Hazeltine, and S. M. Mahajan, Phys. Rev. Lett. {\bf 115}, 025002 (2015).

\bibitem{brady2014}C. S. Brady, C. P. Ridgers, T. D. Arber, and A. R. Bell, Phys. Plasmas {\bf  21}, 033108 (2014).

\bibitem{chang2017}H. X. Chang, B. Qiao, Y. X. Zhang,   Z. Xu, W. P. Yao, C. T. Zhou, and   X. T. He, Phys. Plasmas {\bf 24}, 043111 (2017).

\bibitem{ji2011}L. Ji, B. Shen, X. Zhang, M. Wen, C. Xia, W. Wang, J. Xu, Y. Yu, M. Yu, and Zh. Xu, Phys. Plasmas {\bf 18}, 083104 (2011).

\bibitem{roth2013}M. Roth, D. Jung, K. Falk, N. Guler, O. Deppert, M. Devlin, A. Favalli, J. Fernandez, D. Gautier, M. Geissel {\it et. al.} Phys. Rev. Lett. {\bf 110}, 044802 (2013).

\bibitem{esirkepov2004}T. Esirkepov, M. Borghesi, S. V. Bulanov, G. Mourou and T. Tajima, Phys. Rev. Lett. {\bf 92}, 175003 (2004).

\bibitem{macchi2009}A. Macchi, S. Veghini and F. Pegoraro, Phys. Rev. Lett. {\bf 103}, 085003 (2009); A. Macchi, S. Veghini, T. V. Liseykina and F. Pegoraro, New J. Phys. {\bf 12}, 045013 (2010).
\bibitem{gonoskov2009} A. A. Gonoskov, A. V. Korzhimanov, V. I. Eremin, A. V. Kim, and A. M. Sergeev,  Phys. Rev. Lett. {\bf 102}, 184801 (2009).

\bibitem{yin2011}L. Yin, B. J. Albright, K. J. Bowers, D. Jung, J. C. Fernandez, and B. M. Hegelich, Phys. Rev. Lett. {\bf 107}, 045003 (2011). 
\bibitem{mackenroth2016} F. Mackenroth, A. Gonoskov, M. Marklund, Phys. Rev. Lett. {\bf 117}, 104801 (2016).
\bibitem{matys2019} M. Matys, K. Nishihara, M. Kecova, J. Psikal, G. Korn, and S. V. Bulanov, arXiv:1907.03489 (2019).

\bibitem{macchi2013} A. Macchi, M. Borghesi, and M. Passoni, Rev. Mod. Phys.  {\bf 85}, 751 (2013).
\bibitem{naumova2009} N. Naumova, T. Schlegel, V. T. Tikhonchuk, C. Labaune, I. V. Sokolov and G. Mourou,  Phys. Rev. Lett. {\bf 102}, 025002 (2009).
\bibitem{bulanov2004}S. V. Bulanov, T. Z. Esirkepov, V. S. Khoroshkov, A. V. Kunetsov, and F. Pegoraro, Phys. Lett. A {\bf 299}, 240 (2002).

\bibitem{bulanov2012}S. S. Bulanov, C. B. Schroeder, E. Esarey, and W. P. Leemans, Phys. Plasmas {\bf 19}, 093112 (2012).

\bibitem{tamburini2010}M. Tamburini, F. Pegoraro, A. Di Piazza, C. H. Keitel, and A. Macchi New J. Phys.  \textbf{12}, 123005 (2010).
\bibitem{trapping}A. Gonoskov, A. Bashinov, I. Gonoskov, C. Harvey, A. Ilderton, A. Kim, M. Marklund, G. Mourou, and A. Sergeev, Phys. Rev. Lett. {\bf 113}, 014801 (2014); L. L. Ji, A. Pukhov, I. Yu. Kostyukov, B. F. Shen, and K. Akli, Phys. Rev. Lett. {\bf 112}, 145003 (2014); A. M. Fedotov, N. V. Elkina, E. G. Gelfer, N. B. Narozhny, and H. Ruhl, Phys. Rev. A {\bf 90}, 053847 (2014).
\bibitem{longfield}E. Gelfer, N. Elkina, A. Fedotov, Sci. Rep. \textbf{8}, 6478  (2018); E. G. Gelfer, A. M. Fedotov, S. Weber, PPCF \textbf{60}, 064005  (2018).
\bibitem{dipiazza2012} A. Di Piazza, C. Muller, K. Z. Hatsagortsyan, and C. H. Keitel,  Rev. Mod. Phys. \textbf{84}, 1177 (2012).

\bibitem{ELI} http://www.eli-beams.eu
\bibitem{ELINP} http://www.eli-np.ro/
\bibitem{Apollon} http://portail.polytechnique.edu/luli/en/cilex-apollon/apollon
\bibitem{guo2018}Z. Guo,  L. Yu, J. Wang, C. Wang, Y. Liu, Z. Gan, W. Li, Y. Leng, X. Liang, R. Li, Optics Express,  {\bf 26}, 26776--26786 (2018).

\bibitem{chen}F. Chen, {\it Introduction to Plasma Physics and Controlled Fusion} (Springer US, 1984) Volume 1: Plasma Physics.

\bibitem{ap1956}A. I. Akhiezer and R. V. Polovin, Sov. Phys. JETP \textbf{3}, 696--705 (1956).
\bibitem{kaw1970}P. Kaw and J. Dawson, Phys. Fluids \textbf{13}, 472 (1970).

\bibitem{cattani2000}F. Cattani, A. Kim, D. Anderson, and M. Lisak, Phys. Rev. E {\bf 62}, 1234 (2000).
\bibitem{goloviznin2000} V. V. Goloviznin and T. J. Schep, Phys. Plasmas {\bf 7}, 1564 (2000).
\bibitem{weng2012}S. M. Weng, P. Mulser, and Z. M. Sheng, Phys. Plasmas {\bf 19}, 022705 (2012); S. M. Weng, M. Murakami, P. Mulser, and Z. M. Sheng, New Journal of Physics {\bf 14}, 063026  (2012).
\bibitem{siminos2012}E. Siminos, M. Grech, S. Skupin, T. Schlegel, and V. T. Tikhonchuk Phys. Rev. E {\bf 86}, 056404 (2012).
\bibitem{siminos2017}E. Siminos, M. Grech, B. Svedung Wettervik, T F{\"u}l{\"o}p, New J. Phys.  {\bf 19} 123042 (2017).
\bibitem{ji2018}L. Ji, B. Shen, X. Zhang, New J. Phys. {\bf 20}, 053043 (2018).

\bibitem{vshivkov1998}V. A. Vshivkov, N. M. Naumova, F. Pegoraro, and S. V. Bulanov, Phys. Plasmas {\bf 5}, 2727 (1998).

\bibitem{SM} Supplemental material [URL to be inserted by publisher] illustrating the laser-target evolution.

\bibitem{macchi2005}A. Macchi, F. Cattani, T. V. Liseykina, F. Cornolti, Phys. Rev. Lett. {\bf 94}, 165003 (2005).
\bibitem{klimo2008}O. Klimo, J. Psikal, J. Limpouch, V.T. Tikhonchuk, Phys. Rev. STAB {\bf 11}, 031301 (2008).

\bibitem{pegoraro_prl2007}F. Pegoraro and S. V. Bulanov, Phys. Rev. Lett \textbf{99}, 065002 (2007).


\bibitem{wang2011}H. Y. Wang, C. Lin, Z. M. Sheng, B. Liu, S. Zhao, Z. Y. Guo, Y. R. Lu, X. T. He, J. E. Chen, and X. Q. Yan, Phys. Rev. Lett. {\bf 107}, 265002 (2011). 
\bibitem{wei2017}W. Q. Wei, X. H. Yuan, Y. Fang, Z. Y. Ge, X. L. Ge, S. Yang, Y. F. Li, G. Q. Liao, Z. Zhang, F. Liu {\it et. al.}, Phys. Plasmas {\bf 24}, 113111 (2017).


\bibitem{thaury2007}C. Thaury, F. Quere, J.-P. Geindre, A. Levy, T. Ceccotti, P. Monot, M. Bougeard, F. Reau, P. d’Oliveira, P. Audebert, R. Marjoribanks,  Ph. Martin, Nat. Phys.  {\bf 3}, 424 -- 429 (2007).
\bibitem{behmke2011}M. Behmke, D. an der Br{\"u}gge, C. R{\"o}del, M. Cerchez, D. Hemmers, M. Heyer, O. J{\''a}ckel, M. K{\''u}bel, G. G. Paulus, G. Pretzler {\it et. al.} Phys. Rev. Lett. {\bf 106}, 185002 (2011).

\bibitem{ceccotti2007} T. Ceccotti, A. Levy, H. Popescu, F. Reau, P. D’Oliveira, P. Monot, J. P. Geindre, E. Lefebvre, and Ph. Martin Phys. Rev. Lett. {\bf 99}, 185002 (2007).
\bibitem{green2014}J. S. Green, N. P. Dover, M. Borghesi, C. M. Brenner, F. H. Cameron, D. C. Carroll, P. S. Foster, P. Gallegos, G. Gregori, P. McKenna {\it et. al.}, Plasma Phys. Controlled Fusion {\bf 56}, 084001 (2014).
\bibitem{fang2016}Y. Fang, X. L. Ge, S. Yang, W. Q. Wei, T. P. Yu, F. Liu, M. Chen, J. Q. Liu, X. H. Yuan, Z. M. Sheng, and J. Zhang, Plasma Phys. Controlled Fusion {\bf 58}, 075010 (2016).

\bibitem{robinson2008}A. P. L. Robinson, M. Zepf, S. Kar, R. G. Evans and C. Bellei, New J. Phys. {\bf 10} 013021 (2008).






\bibitem{SMILEI}J. Derouillat, A. Beck, F. Perez, T. Vinci, M. Chiaramello, A. Grassi, M. Fle, G. Bouchard, I. Plotnikov, N. Aunai, J. Dargent, C. Riconda, M. Grech, Comput. Phys. Commun. {\bf 222}, 351--373 (2018).

\end{thebibliography}
\end{document}